\documentclass[conference]{IEEEtran}
\IEEEoverridecommandlockouts
\usepackage{cite}
\usepackage{amsmath,amssymb,amsfonts}
\usepackage{algorithmic}
\usepackage{graphicx}
\usepackage{textcomp}
\usepackage{xcolor}
\usepackage{hyperref}
\usepackage{soul}
\usepackage{flushend}

\begin{document}
\title{Checkpointing SPAdes for Metagenome Assembly: Transparency versus Performance in Production}

\author{\IEEEauthorblockN{Twinkle Jain}
\IEEEauthorblockA{\textit{Northeastern University} \\
jain.t@northeastern.edu}
\and
\IEEEauthorblockN{Jie Wang}
\IEEEauthorblockA{\textit{DOE-JGI Lawrence Berkeley National Laboratory}\\
jwang7@lbl.gov}}


\maketitle

\begin{abstract}
The SPAdes assembler for metagenome assembly is a long-running application commonly used at the NERSC supercomputing site. However, NERSC, like many other sites, has a 48-hour limit on resource allocations.  The solution is to chain together multiple resource allocations in a single run, using checkpoint-restart.  This case study provides insights into the ``pain points'' in applying a well-known checkpointing package (DMTCP: Distributed MultiThreaded CheckPointing) to long-running production workloads of SPAdes.  This work has exposed several bugs and limitations of DMTCP, which were fixed to support the large memory and fragmented intermediate files of SPAdes.  But perhaps more interesting for other applications, this work reveals a tension between the transparency goals of DMTCP and performance concerns due to an I/O bottleneck during the checkpointing process when supporting large memory and many files.  Suggestions are made for overcoming this I/O bottleneck, which provides important ``lessons learned'' for similar applications.
\end{abstract}

\vspace{-2.2mm}
\section{Introduction}
\label{sec:introduction}

Next-generation sequencing (NGS) technology has dramatically changed the scale of genomics study in the past decade. Metagenome sequencing for characterizing the microbial community composition is one of the areas that benefits the most from NGS technology~\cite{venter2004environmental}. The application of metagenome sequencing in microbial community analysis remains challenging --- especially in the area of metagenome assembly~\cite{10.1093/bioinformatics/btr520}.

For metagenome analysis, we use the {\em St. Petersburg genome assembler}, hereafter known simply as SPAdes.
SPAdes has been adopted by the genomics scientific community for a wide variety of metagenomic characterization applications, and has been used for COVID-19 genome assembly~\cite{carbo2020coronavirus,garces2020full}. The original SPAdes assembler paper has been cited over 11,000 times since it was first published in 2012~\cite{bankevich2012spades}.

The SPAdes assembler for metagenome assembly is orchestrated as a pipeline overseen by a Python process, which calls executables (SPAdes binaries) to perform metagenome assembly. The SPAdes uses OpenMP~\cite{dagum1998openmp} to parallelize DNA string K-mer analysis and assembly graph construction, which are both time consuming and computationally intensive. 

Because of the long run time for SPAdes, it was decided to introduce checkpoint-restart based on DMTCP (Distributed MultiThreaded Checkpointing)~\cite{ansel2009dmtcp} for two primary reasons:
\begin{enumerate}
    \item The SPAdes run time on large data sets often exceeds the 48-hour wall-time limit set by the HPC provider's queue policy;
    \item Unexpected job interruptions (e.g., unplanned system shutdowns) can result in days of computation results being lost.
\end{enumerate}

In order to test the efficacy of DMTCP, we use three metagenome datasets (Table \ref{tab:datasets}). The read and base counts imply the scale of the SPAdes assembler. We used small- and medium-scale datasets, namely \textit{Bog} and \textit{Spike-in}, for our initial DMTCP compatibility testing. However, the checkpoint-restart feature is not crucial at the small and medium scales due to the relatively short job run times. Therefore, we focus on the large dataset, \textit{Rhizosphere}, in this case study. All experiments were performed on the National Energy Research Scientific Computing Center's (NERSC's) Cori cluster~\cite{cori} using Intel Skylake nodes.

\vspace{-5.5mm}
\begin{table}[ht]
\centering
\caption{\label{tab:datasets} Datasets used in this study}
\begin{tabular}{|l|c|c|}
\hline
\textbf{Name} & \textbf{Read Count (Millions)} & \textbf{Base Count (Billions)} \\ \hline
\texttt{Bog}           & 31.1          & 4.5       \\ \hline
\texttt{Spike-in}      & 78.7          & 11.8      \\ \hline
\texttt{Rhizosphere}   & 193         & 28.5      \\ \hline
\end{tabular}
\end{table}


The rest of this paper is organized as follows.
Section~\ref{sec:approach} describes our approach to checkpointing, and the issues we faced in applying this approach to complex software at large scale.
Section~\ref{sec:lessons-learned} highlights the lessons learned, in particular the issue of I/O bottlenecks in supporting SPAdes.  DMTCP was chosen for its support for transparent checkpointing, thus easing the burden of the application developer.  While this transparency was successful, it resulted in excessively long checkpointing times. Thus, for the next iteration, we propose modifications to how SPAdes stores its data, and the possible introduction of separate I/O middleware, in order to alleviate the I/O bottleneck during checkpointing.
Section~\ref{sec:conclusion} presents a conclusion.

\section{Approach to Checkpointing}
\label{sec:approach}

In the process of scaling DMTCP for SPAdes, we faced two kinds of challenges: a) fundamental changes needed in DMTCP to make it compatible with SPAdes regardless of the scale; and b) elimination of limitations and bugs exposed by running at larger scales. We discuss the issues faced in detail below.

\subsection{Challenges in making DMTCP compatible for SPAdes}
\subsubsection{Precious files}
One of the fundamental requirements of any checkpoint package is to be able to save enough information to restart the application correctly. One of the key characteristics of the SPAdes application is that it creates/uses temporary files of up to a terabyte in size, known as \emph{precious files}, and later cleans up those temporary files. These files may or may not be opened by the processes at all times.

By default, DMTCP assumes that all the files associated with a process will be persistent and present at their original path. Though DMTCP provides an option to save files opened by each process at checkpoint time, a subset of precious files may not even show up in the opened file list. Note that precious files can be associated in a per-process fashion or be shared among more than one process at the application level. To solve this problem, we exploited DMTCP's plugin-based architecture~\cite{arya2016design} by creating a plugin specifically for SPAdes to handle the precious file issue. This plugin takes a backup of precious files at the checkpoint and replaces them at restore.

\subsubsection{Tuning DMTCP for NERSC environment}
We made minor changes in DMTCP to perform well with the working environment, i.e., the Cori cluster at NERSC. For example, the Cori OS upgrade from CLE6 to CLE7 exposed a memory region merge issue in DMTCP. This edge case was not expected by DMTCP developers. So, we introduced guard pages around the sensitive memory region.

Moreover, Cori was changed in the last year to enable the huge memory pages module by default~\cite{hugepages}. Only after initial debugging did we realize that DMTCP doesn't support huge pages (2~MB pages). So, we disabled the hugepages module explicitly and informed the DMTCP developers about this scenario. The DMTCP developers plan to introduce support for huge pages.

\subsection{Challenges in scaling DMTCP to production workload}

\subsubsection{Data Structure limitations}
Scaling to a real-world scenario required a few modifications in DMTCP's code. For example, for a large dataset, SPAdes uses more file descriptors than the maximum number assumed by DMTCP, causing a clash with DMTCP internal file descriptors. We resolved this error by dynamically choosing the internal file descriptor range for DMTCP to avoid any overlap with the application's file descriptors. Additionally, we increased limits on a few more internal data structures in DMTCP as we switched to production scale from small scale. 

\subsubsection{The ABA bug}
A broad range of large-scale applications have been using DMTCP for checkpoint-restart purposes. However, SPAdes, with a large dataset, exposed the ABA problem~\cite{aba}, a race condition that led to memory corruption in DMTCP. SPAdes uses many user-threads along with OpenMP threads. We observed this memory corruption more often for the large dataset than for the small and medium scale datasets. This memory corruption bug was one of the hardest among other problems to identify because of its non-deterministic nature. Upon reporting the bug to the DMTCP team, we learned that this was a known bug, and they soon provided a fix\footnote{https://github.com/dmtcp/dmtcp/pull/851}.

\subsubsection{Irrecoverable checkpoint state}
SPAdes assembler runs from a Python script. Therefore, at each checkpoint instance, DMTCP needs to save both the Python and SPAdes binary processes. The Python process maintains a relatively small memory footprint (a few MBs) at all times. 
With large datasets, SPAdes binary's memory footprint can become very large (hundreds of GB) many times during the lifetime of the SPAdes assembler.

Consider the general case where $C_i$ $= \{p_{i1}, p_{i2},..., p_{im}\}$ as the $i^{th}$ checkpoint instance of an application with $m$ processes and $p_{ij}$ where $j \in [1,m]$, is the checkpoint image for the $j^{th}$ process at the $i^{th}$ instance. By default, DMTCP overwrites the all the checkpoint images from the $C_{i}$ instance with new ones from the $C_{i+1}$ checkpoint instance on the persistent storage. 

We observed that all processes of the application running under DMTCP replace their own checkpoint image asynchronously, without waiting for other processes to finish the save task. This sets up the potential for a scenario in which process $P_j$ has already replaced $p_{ij}$ with $p_{{i+1}j}$ but process $P_k$ (where $j \neq k$) has not yet finished the checkpointing task. At this point, if the job hits the wall-time limit or crashes, we will have a set of checkpoint images from different instances on the persistent storage, from which a restore is not possible.

We faced this unrecoverable state many times because of the combination of a small (Python) and a large (SPAdes binary) memory footprint. We resolved this issue by introducing a global barrier in DMTCP that allows each process to replace the older checkpoint image synchronously after all processes finish their checkpoint task. 


\section{Lessons Learned:  The Tension between Transparency and Performance}
\label{sec:lessons-learned}
DMTCP is a transparent checkpointing package.  Its long-term philosophy is that if an application runs well natively, then DMTCP should also execute it well.  No change to the target application is needed.  Here in this paper, we are seeing the limits to this philosophy of transparency.  Transparency fails in the following ways:

(i)~The precious files had to be declared explicitly to a new, specialized DMTCP plugin, thus partially breaking the transparency;

(ii) The use of DMTCP with SPAdes introduced a new requirement to write out the precious files. The precious files were part of a database of many files.  Checkpointing required DMTCP to save each such file sequentially. But the native execution of SPAdes does not have to write out the precious files. This is a new application requirement added only because of DMTCP. As a consequence, checkpoint performance suffers greatly, especially because DMTCP performs na\"ive I/O and does {\em not} use an I/O middleware library~\cite{xie2012characterizing, zheng2013flexio} to optimize I/O.  So, DMTCP can no longer easily hide itself and remain transparent.

\begin{figure}[ht!]
  \centering
  \begin{minipage}{0.48\textwidth}
    \centering
    \includegraphics[width=\textwidth]{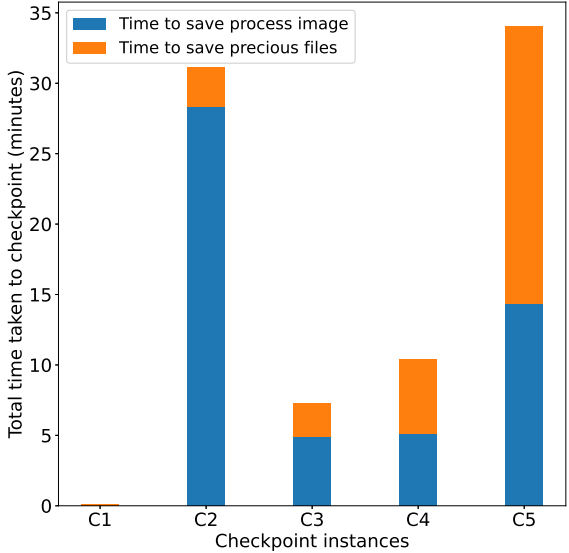}
  \caption{\label{fig:ckpt-time}
           Checkpoint time distribution: SPAdes runs under DMTCP with five periodic checkpoints demonstrating the variability in the time to save precious files (orange) and process images (blue) at each instance.}
\vspace{-1.0mm}
  \end{minipage}
\end{figure}

(iii)~ DMTCP has no feedback from SPAdes about which output file is a precious file and which one is not.  Because it cannot identify precious files, DMTCP ends up saving all the output files naively. This increases the size of the precious files over time (see Figure~\ref{fig:ckpt-time}). It can reach up to a few terabytes, which hinders performance.
\vspace{-4.5mm}

One could argue that, since our primary motivation is to overcome job wall-time limits, we can eliminate this I/O bottleneck while remaining almost transparent by checkpointing only once near the end of the scheduled time, and exit immediately after the checkpoint task completes. This way all the precious files will remain on the persistent storage and no backup would be required.
However, this proposed solution has two problems:

a)~This solution requires an accurate prediction of checkpoint time duration to ensure the completion of checkpoint task before the job wall-time runs out. The memory footprint of the SPAdes application depends on the analysis step that the assembler is performing at the time of the checkpoint. We have observed that the memory footprint varies from a few megabytes to hundreds of gigabytes inconsistently during the lifetime of the SPAdes assembler.

b)~ The Cori Burst Buffer is not available on the Skylake nodes. So DMTCP must use Cori's shared Lustre file system as persistent storage to write checkpoint images. However, the I/O performance of the shared Lustre file system can significantly decrease in the presence of congestion~\cite{nerscfile} at the NERSC supercomputing site. Therefore, there is no straightforward way to predict checkpoint completion time in this setting.

\textbf{Less transparency, better performance:}
The emphasis on transparency clearly caused performance degradation. Therefore, for reasons of performance, it is important to include some efforts on the application side. Modifying applications to improve the I/O is an age-old problem~\cite{xie2012characterizing, zheng2013flexio}.  Nevertheless, some relatively non-invasive changes on the side of SPAdes can reduce the I/O bottleneck. We list three suggestions for SPAdes to improve checkpointing performance:
 
\begin{enumerate}
  \item SPAdes could help DMTCP identify the precious files among regular output files. This could be done with: i)~a simple prefix/suffix to tag temporary/precious files; ii)~writing all precious files in a specific directory which can be saved at checkpoint.
  \item SPAdes can be made checkpoint-aware by introducing a simple ``ckpt-enable'' option. We know that precious files are those temporary files that get deleted at some point in the lifetime of SPAdes. So, when the checkpoint option is enabled then SPAdes don't remove those temporary files from persistent storage.
  \item SPAdes can trigger DMTCP's application-initiated checkpoint. So, instead of initiating checkpoints periodically, initiate a checkpoint when either the memory footprint of SPAdes or the total size of the precious files is smallest.
\end{enumerate}

Any of the above suggestion should reduce the current checkpoint overhead introduced by precious files. We believe that an understanding of both the application (SPAdes) and DMTCP, together, is needed to provide good performance.

\section{Conclusion}
\label{sec:conclusion}

DMTCP's checkpoint-restart has been demonstrated to work well with SPAdes. Numerous challenges were experienced and resolved during testing of DMTCP checkpoint-restart primarily on large metagenome assembly jobs. At the same time, this exposed bugs in DMTCP when handling jobs with large memory footprints. Overall, DMTCP can be extended through minor modifications to work with real-world applications.  The checkpoint overhead was found high due to an added support to handle numerous precious files in SPAdes.
In the future, if these essential precious files could be identified, with the help of the SPAdes developer team, then I/O overhead in periodic checkpoints could be greatly reduced.

\section*{Acknowledgments}
We would like to thank: Zhengji Zhao and Rebecca Hartman--Baker for access to NERSC computing resource, for utility scripts~\cite{VTJ} to re-queue jobs, and for generous assistance in adapting DMTCP; Alicia Clum and Alex Copeland for sequencing datasets and for help with SPAdes; and Gene Cooperman for guidance on debugging using DMTCP with the SPAdes project.
We also thank the reviewers for valuable comments that greatly improved the exposition.
This research used resources of the National Energy Research Scientific Computing Center, which is supported by the Office of Science of the U.S. Department of Energy under Contract No. DE-AC02-05CH11231.
This work was partially supported by National Science Foundation Grant OAC-1740218 and a grant from Intel Corporation.
\IEEEtriggeratref{6}
\bibliographystyle{IEEEtran}
\bibliography{supercheck-spades}

\end{document}